# Multifractals of Central Place Systems: Models, Dimension Spectrums, and Empirical Analysis

Yanguang Chen

Department of Geography, Peking University, Beijing 100871, PRC. E-mail: chenyg@pku.edu.cn

**Abstract**: Central place systems have been demonstrated to possess self-similar patterns in both the theoretical and empirical perspectives. A central place fractal can be treated as a monofractal with a single scaling process. However, in the real world, a system of human settlements is a complex network with multi-scaling processes. The simple fractal central place models are not enough to interpret the spatial patterns and evolutive processes of urban systems. It is necessary to construct multi-scaling fractal models of urban places. Based on the postulate of intermittent space filling, two typical multifractal models of central places are proposed in this paper. One model is put forward to reflect the process of spatial convergence (aggregation), and the generalized correlation dimension varies from 0.7306 to 1.3181; the other model is presented to describe the process of spatial divergence (diffusion), the generalized correlation dimension ranges from 1.6523 to 1.7118. As a case study, an analogy is drawn between the theoretical models and a system of cities and towns of Central Plains, China. A finding is that urban systems take on multifractal form, and can be modeled with multi-scaling fractals. This is a preliminary attempt to develop the theory of fractal central places, and the results are helpful for understanding the similarities and differences between the geographical process of spatial aggregation and that of spatial diffusion.

# 1. Introduction

Fractal geometry represents a new paradigm for geographical studies, and in recent years, the impact of fractal ideas upon geography becomes more significant than ever. The most important theory in human geography is the models of central places, which seeks to explain the size, number, and location of human settlements in a regional system (Goodall, 1987). Central place theory was created by Christaller (1933/1966) and consolidated by Lösch (1945/1954). A central place is a settlement or a nodal point of transport network that, by its functions, serves an area round about it for goods and services (Mayhew, 1997). The spatial texture of a central place network was theoretically proved to be fractal (Arlinghaus, 1985; Arlinghaus and Arlinghaus, 1989; Lam and De Cola, 1993), while the spatial structure of a central place system was empirically demonstrated to be of self-similarity (Chen and Zhou, 2006). The number, size, and distance between different locations of human settlements can be formulated as three exponential laws, from which it follows three power laws indicative of allometric relations and fractals (Chen, 2011). Central place fractals belong to fractal cities (Batty and Longley, 1994). However, the existing models of central place fractals are mainly simple fractals (monofractals), which are based on one scaling process. The real settlement systems are very complex and should be described with multifractals, which are based on multi-scaling processes. Frankhauser (1998) proposed a multifractal model of transport network in terms of central place patterns, but the research remains to be developed. Actually, it is necessary to apply the geometry of multifractals to central place systems to advance the theory and method of spatial analysis (Frankhauser, 2008).

There are two approaches to studying central place fractals. One is the theoretical approach, and the other, the empirical approach. The former starts from model building, while the latter begins from data analysis. In fact, any mathematical tool has two functions for scientific researches: one is to process experimental or observational data, and the other is to construct postulates, make models, and develop theory. If the fractal concept is employed to build mathematical models for cities or systems of cities, it is a theoretical approach; if the fractal geometry is used as a means of data processing for urban systems, it is an empirical approach. In geography, generally speaking, the fractal-based theoretical approach comprises the following steps: defining postulated conditions, proposing/deriving fractal models, and, if possible, testifying/verifying the models, revealing laws, rules or principles, and the aim is to propound a theory. The empirical approach consists of the below steps: defining a study area, obtaining observed data, computing fractal dimensions, making an analysis with the fractal parameters, and the aim is to bring to light the temporal processes and spatial patterns of geographical systems. In some cases, the two approaches can be utilized in the same study to supplement one another.

Multifractals theory has been applied to geographical and urban studies, and the results are interesting and revealing (Appleby, 1996; Ariza-Villaverde et al, 2013; Cavailhès *et al*, 2010; Chen and Zhou, 2004; Frankhauser, 2008; Haag, 1994). In this paper, the theoretical approach is employed to research central place fractals. Two multi-scaling fractal models of central places will be presented. One is to reflect spatial concentration of central places, and the other is to mirror spatial deconcentration of cities and towns. The rest parts are organized as follows. In Section 2, theoretical postulates, geometrical models, and dimension spectrums will be presented. Then, in Section 3, an analogy will be drawn between the fractal models and the real patterns of human settlements in Central Plain, China. A number of questions will be discussed in Section 4. The main significance of this study is that new models of multifractals are constructed for urban studies, and the multifractals are revealing for our understanding the geographical processes and patterns of human settlements.

## 2. Multifractal models

### 2.1 Basic postulates and main parameters

The classical models of central places are based on the growth processes of equal probability and intermittency-free space filling. Therefore, the settlements in the same class are evenly distributed over an area. Despite the self-similar texture consisting of fractal lines (Arlinghaus, 1985), the theoretical dimension of spatial structure of central place systems is $d$=2 (Chen, 2011). However, the actual patterns of cities and towns are not spatially homogeneous, and the empirical dimension is a fraction rather an integer (Chen and Zhou, 2006). In other words, the central place patterns are fractals in the real world. In order to present the multifractal models of central places, two postulates are given as follows. First, the space filling of central place development is a process of intermittency. Second, the growth of human settlements is not of equal probability.

The human geographical processes can be distributed into three classes: spatial concentration, spatial deconcentration, and spatial equalization. The process of homogenization leads to a Euclidean space rather than fractal space. The multifractal models of central places are based on the spatial processes of concentration and deconcentration. The process of concentration results in spatial convergence, while the process of deconcentration brings on spatial divergence. Spatial equalization of urban growth is one of the preconditions for the standard patterns of central places. However, the real patterns of human settlements are associated with spatial heterogeneity rather than regional homogeneity. Both the processes of spatial concentration and spatial deconcentration can result in some "transformative patterns" of central place networks (Isard, 1972; Northam, 1979; Saey, 1973). The changed structure of central place systems should be modeled with multifractals instead of monofractals.

Generally speaking, two sets of parameters can be used to characterize a multifractal pattern. One is the *global* parameters, and the other is the *local* parameters. The global parameters comprise the generalized correlation dimension and the mass exponent, while the local parameters consist of the Lipschitz-Hölder exponent, and the fractal dimension of the set supporting this exponent. According to multifractal theory, we have a set of equations and formulae for these parameters (Grassberger, 1983; Grassberger, 1985; Grassberger and Procassia, 1983; Halsey *et al*, 1986; Hentschel and Procaccia, 1983; Meakin, 1998; Stanley and Meakin, 1988; Vicsek, 1989; Vicsek, 1990). The mass exponent and the generalized correlation dimension can be given by the following transcendental equation

$$\sum_{i=1}^{N(r)} P_i^q r_i^{-\tau(q)} = \sum_{i=1}^{N(r)} P_i^q r_i^{(1-q)D_q} = 1, \qquad (1)$$

where $N(r)$ denotes the number of fractal copies, $q$ is the order of moment, $\tau(q)$ refers to the mass exponent of order $q$, and $D_q$ to the generalized correlation dimension. The relationship between $\tau(q)$ and $D_q$ is as below:

$$\tau(q) = (q-1)D_q. \qquad (2)$$

The two parameters $D_q$ and $\tau(q)$ compose the set of global parameters of multifractals sets.

A multifractals set (the whole) has many fractal subsets (parts), and each fractal subset has its local fractal dimension. Each fractal subset has a probability of growth, which follow a power law such as

$$P_i(r_i) = r_i^{\alpha(q)}, \tag{3}$$

where $\alpha(q)$ denotes the Lipschitz-Hölder singularity exponent (Feder, 1988). For given $q$ value, if equation (3) comes into existence, the corresponding local fractal dimension can be given by

$$N_i(\alpha, r_i) = r_i^{-f(\alpha)}, \tag{4}$$

where $N(\alpha, \varepsilon)$ refers to the number of the smaller fractal copies, and $f(\alpha)$ to the fractal dimension of the subset supporting the exponent $\alpha(q)$ (Frisch and Parisi, 1985; Feder, 1988). The parameters $\alpha(q)$ and $f(\alpha)$ constitute the set of local parameters of multifractals sets. In theory, the global parameters and local parameters can be turned into one another by Legendre's transform (Badii and Politi, 1997; Feder, 1988). The equations of Legendre's transform are as follows

$$\alpha(q) = \frac{d\tau(q)}{dq} = D_q + (q-1)\frac{dD_q}{dq}, \tag{5}$$

$$f(\alpha) = q\alpha(q) - \tau(q) = q\alpha(q) - (q-1)D_q. \tag{6}$$

Using these equations, we can translate the global parameters $\tau(q)$ and $D_q$ and into the local parameters $\alpha(q)$ and $f(\alpha)$, and *vice versa*.

**2.2 Multifractal model of spatial concentration**

The first regular multifractal central place model is on spatial concentration, which indicates that a great number of cities and towns grow around a large central city and many subcentral cities in virtue of "centripetal force". In this instance, the spatial pull of a central city is greater than the push of the city. This multifractal model is constructed as follows. The initiator is one cell, and the linear scale of the cell is one unit; the generator comprises 7 consisting of 13 cells, and its linear size is 7 units (Figure 1). The initiator is a special case. Where the whole fractal object is concerned, the number ratio of the fractal generating process is 13, and linear scale ratio is 7. Thus the similar dimension is $D=\ln(13)/\ln(7)=1.3181$, which is in fact the upper limit of the generalized correlation dimension, i.e., $D_{-\infty}= 1.3181$. The good angle of view of examining a multifractal object is the probability of growth and the corresponding linear scale. Comparing the initiator and the generator shows that there are two types of growing probability. For the central part, the linear scale of the fractal copy is $r_0=3/7$, while its growing probability is $p_0=7/13$; For the 6 parts on the periphery, the linear size of the fractal copies is $r_1=r_2=r_3=r_4=r_5=r_6=1/7$, while the growing probability is $p_1=p_2=p_3=p_4=p_5=p_6=1/13$ (Table 1).

Table 1 The numbers and linear scales of fractal copies of multifractal central place model indicative of spatial concentration (the first four steps)

| Step | Number of fractal copies | | | Linear scale | | |
| --- | --- | --- | --- | --- | --- | --- |
| | Center | Periphery | The whole | Center | Periphery | The whole |
| Step 1 (Initiator) | 1 | 6*0 | 1 | 1 | 0 | 1 |
| Step 2 (Generator) | 7 | 6*1 | 13 | 3 | 1 | 7 |
| Step 3 | 91 | 6*7 | 169 | 21 | 7 | 49 |
| Step 4 | 1183 | 6*169 | 2197 | 147 | 49 | 343 |

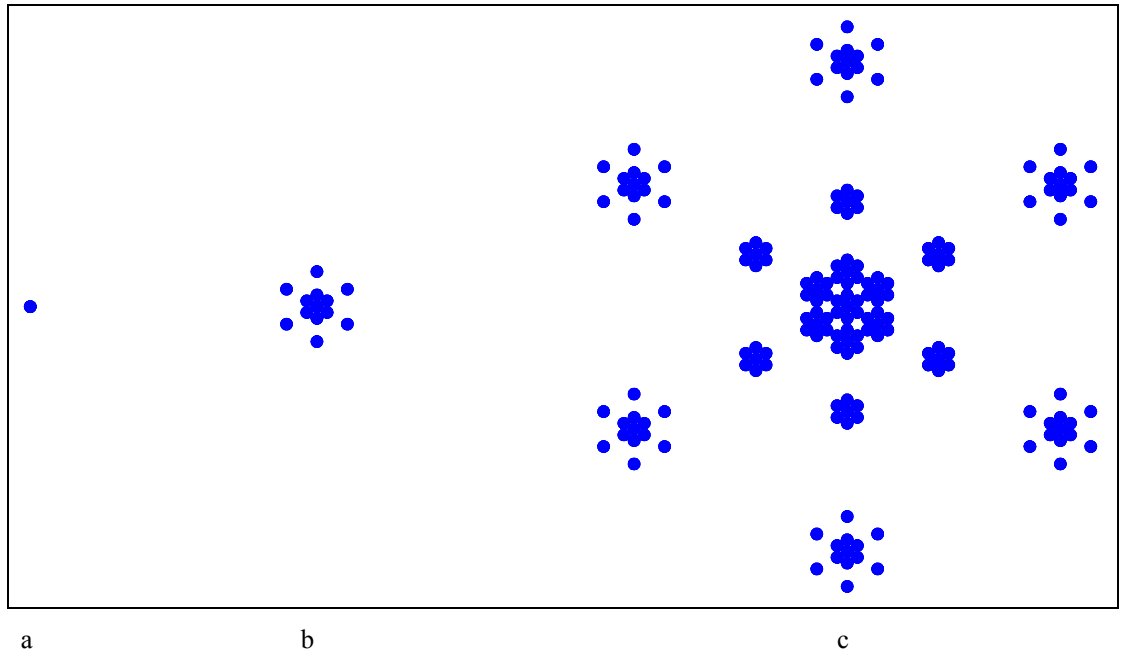

a         b         c

Figure 1 A regular multifractal model of central place systems indicative of spatial concentration

(the first three steps)

Since this is regular fractal object, the global parameters and local parameters can be expressed with determinate formulae. According to equation (1), we have

$$P_0^q r_0^{-\tau(q)} + 6P_1^q r_1^{-\tau(q)} = (\frac{7}{13})^q (\frac{3}{7})^{-\tau(q)} + 6(\frac{1}{13})^q (\frac{1}{7})^{-\tau(q)} = 1. \qquad (7)$$

The generalized correlation dimension will be computed with the follow formula

$$D_q = \begin{cases} \dfrac{P_0 \ln P_0 + 6P_1 \ln P_1}{P_0 \ln r_0 + 6P_1 \ln r_1} = \dfrac{(\frac{7}{13})\ln(\frac{7}{13}) + 6(\frac{1}{13})\ln(\frac{1}{13})}{(\frac{7}{13})\ln(\frac{3}{7}) + 6(\frac{1}{13})\ln(\frac{1}{7})}, & q=1 \\[2ex] \dfrac{\tau(q)}{q-1}, & q \neq 1 \end{cases} \qquad (8)$$

By equation (8), the information dimension is about $D_1 \approx 1.1202$. The Lipschitz-Hölder singularity exponent can be obtained by differentiating equation (7) with respect to $q$, and the result is

$$r_0^{-\tau(q)} P_0^q \ln P_0 - P_0^q r_0^{-\tau(q)} \ln r_0 \frac{d\tau(q)}{dq} + 6r_1^{-\tau(q)} P_1^q \ln P_1 - 6P_1^q r_1^{-\tau(q)} \ln r_1 \frac{d\tau(q)}{dq} = 0. \qquad (9)$$

In terms of equation (5), rearranging equation (9) yields

$$\alpha(q) = \frac{d\tau(q)}{dq} = \frac{r_0^{-\tau(q)} P_0^q \ln P_0 + 6r_1^{-\tau(q)} P_1^q \ln P_1}{r_0^{-\tau(q)} P_0^q \ln r_0 + 6r_1^{-\tau(q)} P_1^q \ln r_1}$$
$$= \frac{(\frac{3}{7})^{-\tau(q)}(\frac{7}{13})^q \ln(\frac{7}{13}) + 6(\frac{1}{7})^{-\tau(q)}(\frac{1}{13})^q \ln(\frac{1}{13})}{(\frac{3}{7})^{-\tau(q)}(\frac{7}{13})^q \ln(\frac{3}{7}) + 6(\frac{1}{7})^{-\tau(q)}(\frac{1}{13})^q \ln(\frac{1}{7})} \cdot \qquad (10)$$

Finally, through the Legendre transform, equation (6), the local fractal dimension can be derived as below

$$f(\alpha) = q\left[\frac{(\frac{3}{7})^{-\tau(q)}(\frac{7}{13})^q \ln(\frac{7}{13}) + 6(\frac{1}{7})^{-\tau(q)}(\frac{1}{13})^q \ln(\frac{1}{13})}{(\frac{3}{7})^{-\tau(q)}(\frac{7}{13})^q \ln(\frac{3}{7}) + 6(\frac{1}{7})^{-\tau(q)}(\frac{1}{13})^q \ln(\frac{1}{7})}\right] - (q-1)D_q. \qquad (11)$$

Several special fractal dimension values can be directly obtained through the formulae from the above equations. The upper limit and lower limit of the generalized correlation are as below

$$D_{-\infty} = \frac{\ln(1/13)}{\ln(1/7)} = 1.3181, \quad D_{\infty} = \frac{\ln(7/13)}{\ln(3/7)} = 0.7306.$$

If we use the box-counting method to estimate the fractal dimension, the result will be $D_{-\infty} = \ln(13)/\ln(7) \approx 1.3181$ instead of $D_0 \approx 1.1613$. It is easy to prove that $D_0 = \max(f(\alpha))$, that is, the capacity dimension is just the maximum value of the local dimension. In fact, if and only if $q=0$, we will have

$$\frac{df(\alpha)}{dq} = q\frac{d\alpha(\alpha)}{dq} = 0. \qquad (12)$$

which is just the condition of maximum value of the local fractal dimension. For $q=1$, according to equation (2), $\tau(1)=0$, thus, according to equations (2), (10), and (11), $\alpha(1)=f(\alpha)=D_1\approx1.1202$ (Table 2).

Table 2 Partial values of global and local parameters of the multifractal central place model indicative of spatial concentration

| Order of moment | Global parameter | | Local parameter | |
| --- | --- | --- | --- | --- |
| $q$ | $D_q$ | $\tau(q)$ | $\alpha(q)$ | $f(\alpha(q))$ |
| $-\infty$ | **1.3181** | $-\infty$ | **1.3181** | **0.9208** |
| -100 | 1.3142 | -132.7331 | 1.3181 | 0.9208 |
| -2 | 1.2150 | -3.6449 | 1.2740 | 1.0969 |
| -1 | 1.1921 | -2.3842 | 1.2449 | 1.1393 |
| 0 | 1.1613 | -1.1613 | 1.1969 | 1.1613 |
| 1 | 1.1202 | 0.0000 | 1.1202 | 1.1202 |
| 2 | 1.0677 | 1.0677 | 1.0103 | 0.9529 |
| 100 | 0.7380 | 73.0604 | 0.7306 | 0.0000 |
| $\infty$ | **0.7306** | $\infty$ | **0.7306** | **0** |

## 2.3 Multifractal model of spatial deconcentration

Differing from the first model, the second regular multifractal central place model is on spatial deconcentration, which indicates that a large number of cities and towns are distributed around central and subcentral cities on account of "centrifugal force". In this case, the spatial pull of a central city is less than the push of the city. This model is constructed as below. The initiator is one cell, and the linear scale of the cell is one unit. The generator includes 7 parts comprising 43 cells, and its linear size is 9 units (Figure 2). The initiator as a special case is not taken into account. The number ratio of the fractal hierarchy is 43, and the linear scale ratio is 9. In this instance, the similar dimension is $D=\ln(43)/\ln(9)=1.7118$, which suggests $D_{-\infty}=1.3181$. In the process of multifractal development, there are two types of growing probability. For the central part, the linear scale of the fractal copy is $r_0=1/9$, while its growing probability is $p_0=1/43$. For the 6 parts on the periphery, the linear size of the fractal copies is $r_1=r_2=r_3=r_4=r_5=r_6=1/3$, while the growing probability is $p_1=p_2=p_3=p_4=p_5=p_6=7/43$ (Table 3).

Table 3 The numbers and linear scales of fractal copies of multifractal central place model indicative of spatial deconcentration (the first four steps)

| Step | Number of fractal copies | | | Linear scale | | |
|---|---|---|---|---|---|---|
| | Center | Periphery | The whole | Center | Periphery | The whole |
| Step 1(Initiator) | 1 | 0 | 1 | 1 | 0 | 1 |
| Step 2(Generator) | 1 | 6*7 | 43 | 1 | 3 | 9 |
| Step 3 | 43 | 6*301 | 1849 | 9 | 27 | 81 |
| Step 4 | 1849 | 6*12943 | 79507 | 81 | 243 | 729 |

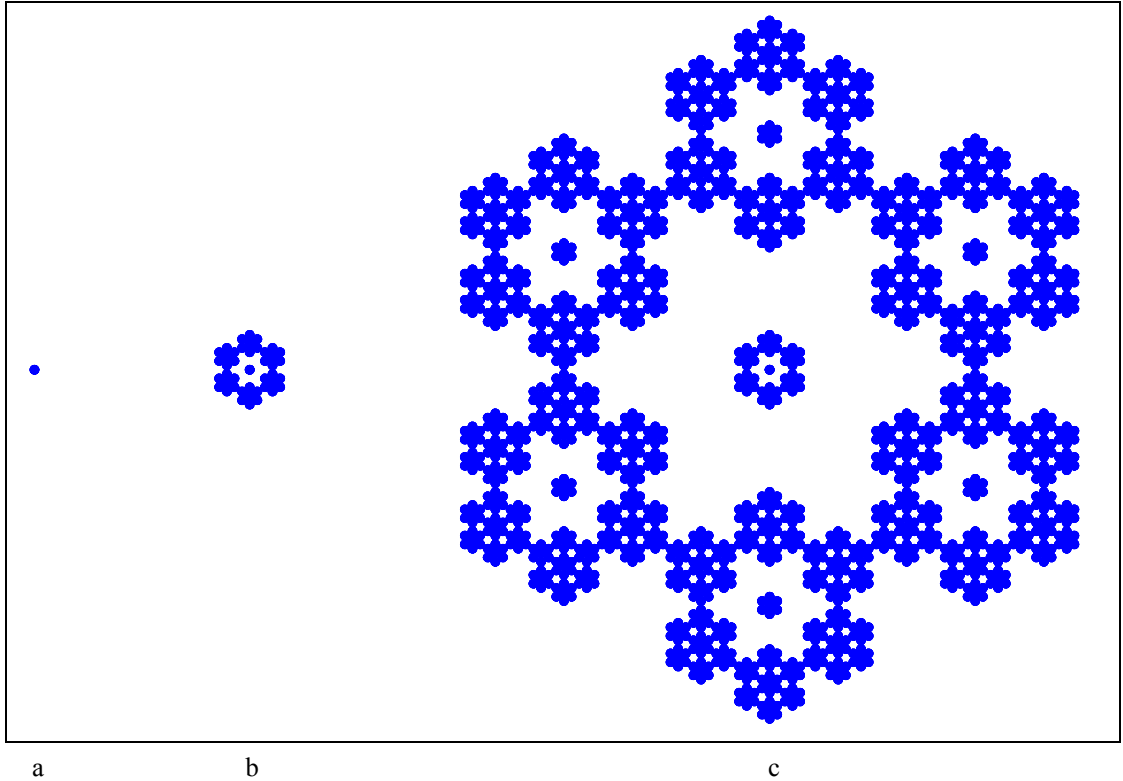

a　　　　　　　　b　　　　　　　　　　　　　　c

**Figure 2 A regular multifractal model of central place systems indicative of spatial deconcentration (the first three steps)**

Revising the mathematical expressions of the first multifractal model displayed in Figure 1, we can obtain the formulae for the second model shown in Figure 2. In terms of equation (1), the mass exponent can be given by

$$P_1^q r_1^{-\tau(q)} + 6 P_2^q r_2^{-\tau(q)} = (\frac{1}{43})^q (\frac{1}{9})^{-\tau(q)} + 6(\frac{7}{43})^q (\frac{3}{9})^{-\tau(q)} = 1. \tag{13}$$

The generalized correlation dimension is defined as

$$D_q = \begin{cases} \dfrac{p_1 \ln p_1 + 6 p_2 \ln p_2}{p_1 \ln r_1 + 6 p_2 \ln r_2} = \dfrac{(\frac{1}{43})\ln(\frac{1}{43}) + 6(\frac{7}{43})\ln(\frac{7}{43})}{(\frac{1}{43})\ln(\frac{1}{9}) + 6(\frac{7}{43})\ln(\frac{1}{3})}, & q = 1 \\ \dfrac{\tau(q)}{q-1}, & q \neq 1 \end{cases} \quad (14)$$

Thus the information is about $D_1$=1.6551. From equation (13), it follows the singularity exponent in the form

$$\alpha(q) = \dfrac{d\tau(q)}{dq} = \dfrac{r_1^{-\tau(q)} P_1^q \ln P_1 + 6 r_2^{-\tau(q)} P_2^q \ln P_2}{r_1^{-\tau(q)} P_1^q \ln r_1 + 6 r_2^{-\tau(q)} P_2^q \ln r_2}$$
$$= \dfrac{(\frac{1}{9})^{-\tau(q)}(\frac{1}{43})^q \ln(\frac{1}{43}) + 6(\frac{1}{3})^{-\tau(q)}(\frac{7}{43})^q \ln(\frac{7}{43})}{(\frac{1}{9})^{-\tau(q)}(\frac{1}{43})^q \ln(\frac{1}{9}) + 6(\frac{1}{3})^{-\tau(q)}(\frac{7}{43})^q \ln(\frac{1}{3})} \quad (15)$$

Accordingly, the local fractal dimension of the set supporting the singularity exponent is

$$f(\alpha) = q\left[\dfrac{(\frac{1}{9})^{-\tau(q)}(\frac{1}{43})^q \ln(\frac{1}{43}) + 6(\frac{1}{3})^{-\tau(q)}(\frac{7}{43})^q \ln(\frac{7}{43})}{(\frac{1}{9})^{-\tau(q)}(\frac{1}{43})^q \ln(\frac{1}{9}) + 6(\frac{1}{3})^{-\tau(q)}(\frac{7}{43})^q \ln(\frac{1}{3})}\right] - (q-1)D_q . \quad (16)$$

Several special fractal parameter values can be directly computed through the above equations. The upper limit and lower limit of the generalized correlation are as follows

$$D_{-\infty} = \dfrac{\ln(1/43)}{\ln(1/9)} = 1.7118, \quad D_{\infty} = \dfrac{\ln(7/43)}{\ln(3/9)} = 1.6523.$$

If we use the box-counting method to estimate the fractal dimension, the result will be $D_{-\infty}$=ln(43)/ln(9)≈1.7118 rather than the capacity dimension $D_0$≈1.6552. For $q$=1, we have $\alpha(1)= f(\alpha)=D_1$≈1.6551 (Table 4).

**Table 4 Partial values of global and local parameters of multifractal central place model indicative of spatial deconcentration**

| Order of moment | Global parameter | | Local parameter | |
|---|---|---|---|---|
| $q$ | $D_q$ | $\tau(q)$ | $\alpha(q)$ | $f(\alpha(q))$ |
| -∞ | **1.7118** | -∞ | **1.7118** | **0** |
| -100 | 1.6949 | -171.1836 | 1.7115 | 0.0300 |
| -2 | 1.6556 | -4.9668 | 1.6562 | 1.6544 |
| -1 | 1.6554 | -3.3108 | 1.6558 | 1.6550 |
| 0 | 1.6552 | -1.6552 | 1.6554 | 1.6552 |
| 1 | 1.6551 | 0.0000 | 1.6551 | 1.6551 |
| 2 | 1.6549 | 1.6549 | 1.6547 | 1.6546 |
| -100 | 1.6949 | -171.1836 | 1.7115 | 0.0300 |
| ∞ | **1.6523** | ∞ | **1.6523** | **1.6309** |

## 2.4 Multifractals dimension spectrums

The real patterns of city distributions in a geographical region are more complex than the theoretical models, but simple models always lead to deep understanding of geographical processes and patterns. For the central place multifractals, there is a great contrast between the first model and the second model presented above. The similarities and differences between the two multifractalses can be reflected by the spectrums of multifractal parameters. The generalized correlation dimension, the mass exponent, the singularity exponent, and the local fractal dimension change with the order of moment $q$. To calculate these parameters, we must solve the transcendental equations (7) and (13). This can be achieved through Matlab, which has a function "*fsolve*". The global and local parameters of the two models as functions of $q$ are displayed in Figures 3 and 4, respectively. The similarities between the two sets of spectral curves are obvious. Comparing Figure 3 with Figure 4 shows the differences between two models. The distinction can be reflected by the global dimension $D_q$ and local dimension $f(\alpha)$--the mass exponent $\tau(q)$ depends on the generalized correlation dimension, while the spectrums of the singularity exponent $\alpha(q)$ is similar to those of the global dimension. First, the correlation dimension spectrum of the first model takes on a form of odd symmetry, but the spectral curve of the second model is not so symmetry. Second, the local fractal dimension spectrum of the first model does not converge if the moment order approaches the negative infinity ($q\to-\infty$), while the spectral curve of the second model does not converge if the moment order approaches positive infinity ($q\to\infty$). For the first model, if $q\to\infty$, we will have $f(\alpha)\to 0$, while if $q\to-\infty$, we will have $f(\alpha)\to\ln(6)/\ln(7)\approx 0.9208$ (Table 3). For the second model, if $q\to-\infty$, we will have $f(\alpha)\to 0$, while if $q\to\infty$, we will have $f(\alpha)\to\ln(6)/\ln(3)\approx 1.6309$ (Table 4).

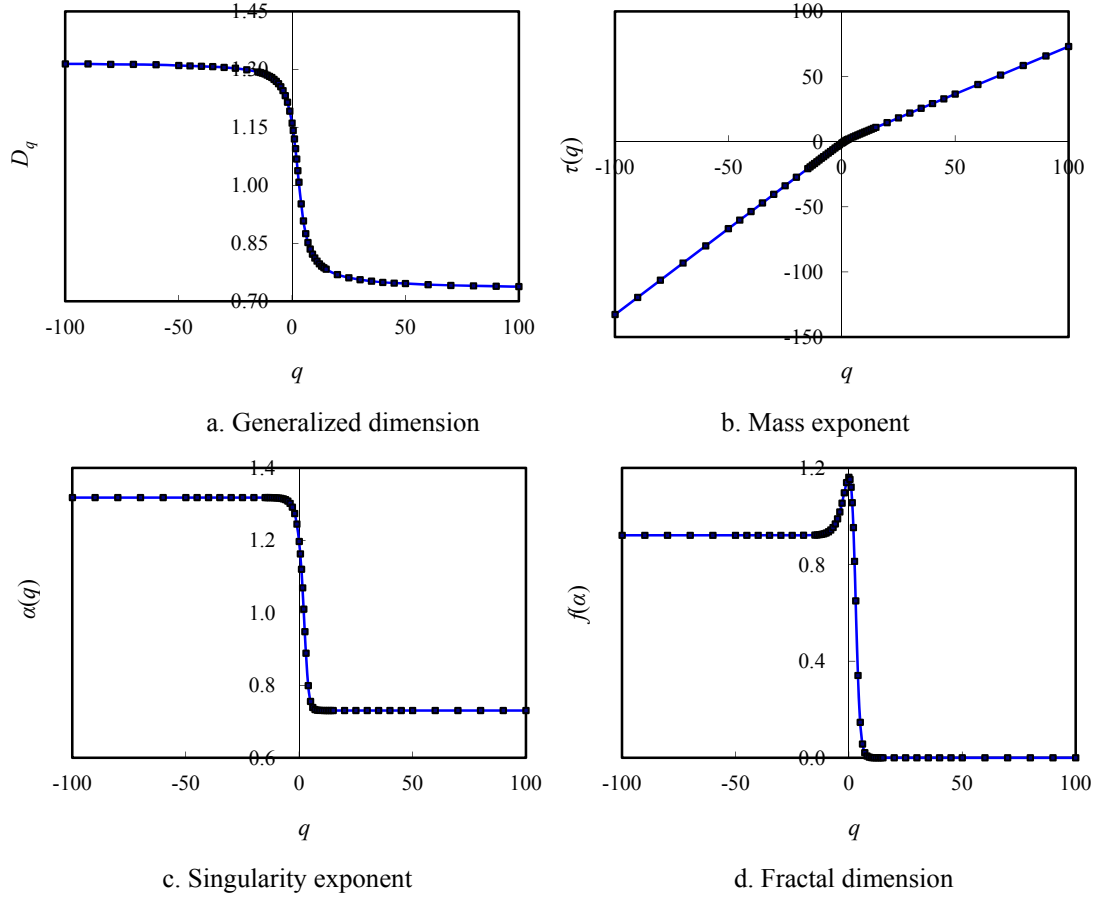

a. Generalized dimension    b. Mass exponent

c. Singularity exponent    d. Fractal dimension

**Figure 3 The spectrum of generalized correlation dimension and the curves of related parameters**

**of the multifractal central place system indicative of spatial concentration**

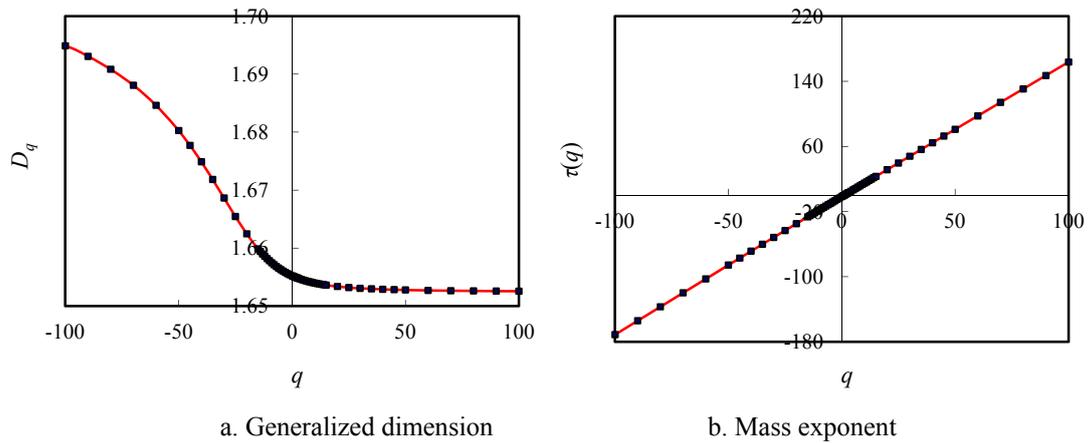

a. Generalized dimension    b. Mass exponent

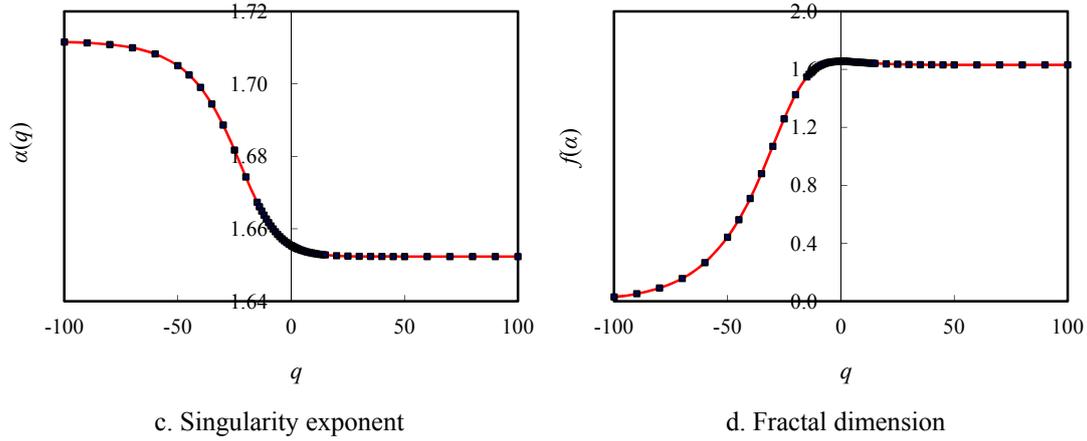

c. Singularity exponent      d. Fractal dimension

**Figure 4 The spectrum of generalized correlation dimension and the curves of related parameters of the multifractal central place system indicative of spatial deconcentration**

The process of urban concentration differs from the process of urban deconcentration. There are sharp contrasts between the multifractal spectrums of the concentration model and those of the deconcentration model (Table 5). The relation of the Lipschitz-Hölder exponent to the fractal dimension of the set supporting this singularity exponent can be illustrated with a plot, which is termed $f(\alpha)$ curve or multifractal spectrum (Carpena et al, 2000; Feder, 1988). This is a kind of unimodal curve. The $f(\alpha)$ curves of the two multifractal models stand in vivid contrast against each other (Figure 5). The unimodel curve of the first model does not converge for upper limit value of the singularity exponent, while the single-peak curve of the second model does not converge for the lower limit value of $\alpha(q)$.

**Table 5 Comparison between the multifractal curves of centralization processes and those of decentralization processes**

| Level | Parameter | Multifractal concentration | Multifractal de concentration |
|---|---|---|---|
| Global | $D_q$ | If $q\to\infty$, $D_q$ converge slowly | If $q\to-\infty$, $D_q$ converge slowly |
|  | $\tau(q)$ | The slopes approach $D_{-\infty}$ and $D_{\infty}$ | The slopes approach $D_{-\infty}$ and $D_{\infty}$ |
| Local | $\alpha(q)$ | The range of singularity is narrow | The range of singularity is wide |
|  | $f(\alpha(q))$ | If $q\to-\infty$, $f(\alpha)$ fails to converge to 0 | If $q\to\infty$, $f(\alpha)$ fails to converge to 0 |

**Note:** For the multifractal concentration model, if $q>1$ and $q\to\infty$, the spatial information of central areas will be revealed; if $q<1$ and $q\to-\infty$, the spatial information of peripheral areas will be disclosed. For the multifractal deconcentration model, if $q>1$ and $q\to\infty$, the spatial information of peripheral areas will be revealed; if $q<1$ and

*q*→-∞, the spatial information of central areas will be disclosed.

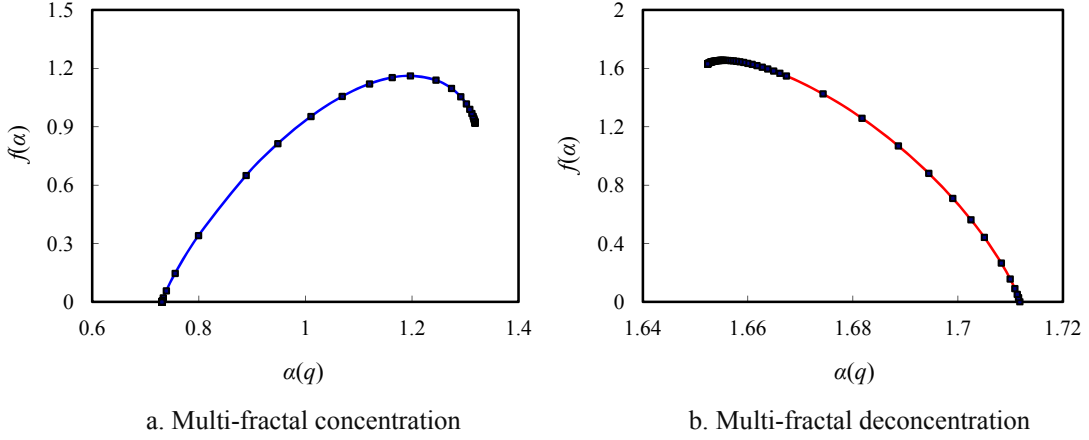

a. Multi-fractal concentration      b. Multi-fractal deconcentration

**Figure 5 The *f(α)* curves of two multifractal processes: spatial concentration and de concentration**

## 3. Empirical analysis

### 3.1 Two approaches to estimating multifractal dimensions

Random fractals in the possible world or the real world differ from regular fractals in the mathematical world. The approach to computing multifractal spectrums of theoretical models is not suitable for random multifractals. In different fields, different methods have been developed and employed to calculate multifractal spectrums (Ariza-Villaverde *et al*, 2013; Carpena *et al*, 2000; Chen and Wang, 2013; Stanley *et al*, 1990; Vicsek, 1990). Generally speaking, to describe an actual mulitfractal object, we need a generalized probability measurement. A real central place system can be treated as a prefractal with multi-scaling processes. The box-counting method can be adopted to characterize this kind of fractal systems (Carpena *et al*, 2000; Chen and Wang, 2013; Vicsek, 1990). Suppose that there is a central place system comprising *n* urban places within a geographical region. We can use a set of boxes, the side length of which is *r*, to cover this region on a digital map. If the number of urban places contained by the *i*th 'box' is $n_i(r)$, then the "probability" of urban places appearing in this box can be defined by the following formula

$$P_i(r) = \frac{n_i(r)}{n} = \frac{n_i(r)}{\sum_{i=1}^{N(r)} n_i(r)}, \qquad (17)$$

in which $N(r)$ denotes the number of nonempty boxes with a linear size of $r$ (the index of nonempty boxes is $i=1,2,3,\ldots$). Based on the probability measurements, the values of the global and local parameters can be estimated by equations (1), (2), (3), and (4) in principle.

However, the approach to computing the of the regular central place multifractals are not completely effectual for the estimating the fractal parameters of the random prefractals such as systems of urban places. In practice, two approaches can be employed to evaluate multifractal spectrums. The first method is to start off with determining the global parameters, and the local parameters can be gain by Legendre's transform. The linear sizes of different fractal copies of regular multifractals are possibly different from one another. However, for random fractals, we do not know the linear sizes of different fractal units at all. Thus, substituting $r$ for $r_i$ in equation (1) yields

$$\sum_{i=1}^{N(r)} P_i(r)^q = r^{(q-1)D_q}, \qquad (18)$$

where $r=r_1=r_2=\ldots$. Equation (18) is suitable for simple regular fractals because that the linear scales of different fractal copies at a given level of a monofractal equal one another. For multifractals, equation (18) is not exact. However, if the linear scale of spatial measurement, $r$, becomes very very small, the estimated value of a fractal parameter, $D_q$, will be very close to its real value. Based on equation (18), the generalized correlation dimension can be defined as

$$D_q = \frac{1}{q-1} \lim_{r \to 0} \frac{\log \sum_{i=1}^{N(r)} P_i(r)^q}{\log r} = -\lim_{r \to 0} \frac{I_q(r)}{\log r}. \qquad (19)$$

where $I_q$ refers to Renyi's entropy. A linear regression equation can be obtained as follow

$$I_q(r) = I_0 - D_q \log(r), \qquad (20)$$

where $I_0$ denotes the initial information entropy of multifractal systems. In theory, $I_0=0$, but empirically, $I_0$ is a near 0. For $q \neq 1$, we have Renyi's information entropy

$$I_q(r) = \frac{1}{1-q} \log \sum_{i=1}^{N(r)} P_i(r)^q; \qquad (21)$$

If $q=1$, according to L'Hospital's rule, the Renyi entropy will change to Shannon's information entropy

$$I_1(r) = -\sum_{i=1}^{N(r)} P_i(r) \log P_i(r). \qquad (22)$$

For multifractal systems of urban places, it is easy to estimate the global parameters. Using the box-counting method, we can calculate the Renyi's or Shannon's information entropy of the spatial distribution of urban places; using linear regression based on equation (19), we can estimate the fractal parameter $D_q$, and $\tau_q=(q-1)D_q$. However, if we want to turn the global parameters into the local parameters with Legendre's transform, we had better calculate a continuous spectrum of $D_q$ or $\tau_q$. It is not often convenient to obtain a continuous spectrum of a global parameter. Thus we need another approach to estimating the multifractal parameters of city fractals. The $\mu$-weight method, proposed by Chhabra and Jensen (1989) and Chhabra et al (1989), can be employed to calculate the values of $\alpha(q)$ and $f(\alpha)$. The global parameters can be gained from the local parameter with Legendre's transform. Defining a weight measurement such as

$$\mu_i(r) = P_i(r)^q / \sum_{i=1}^{N(r)} P_i(r)^q, \qquad (23)$$

we have

$$\alpha(q) = \lim_{\varepsilon \to 0} \frac{1}{\log r} \sum_{i=1}^{N(r)} \mu_i(r) \log P_i(r), \qquad (24)$$

$$f(\alpha) = \lim_{\varepsilon \to 0} \frac{1}{\log r} \sum_{i=1}^{N(r)} \mu_i(r) \log \mu_i(r). \qquad (25)$$

Then taking $\log(r)$ as an independent variable, and $\sum \mu_i(r)\log P_i(r)$ or $\sum \mu_i(r)\log \mu_i(r)$ as a dependent, we can estimate the values of $\alpha(q)$ and $f(\alpha)$ by linear regression technique. Afterwards, the values of $D_q$ and $\tau(q)$ can be determined by the Legendre transform.

### 3.2 A case of China's cities

In the section, the urban places of the Central Plain, China, are employed to testify the models of central place multifractals. The Central Plain (*Zhongyuan*) refers to the region comprising the middle and lower reaches of the Yellow River (*Huanghe*). It includes part of the North China Plain and is regarded as the cradle of Chinese civilization. In the narrowest sense, the Central Plain mainly indicates modern-day Henan Province. As space of the maps of human settlements is limit, the study area cover the whole area of Henan, and the southern part of Hebei, the southern part of Shanxi, the western part of Shandong, and southeastern part of Shaanxi, northern part of Hubei, and northwestern part of Anhui (Figure 6). The systems of human settlements in the Central Plain have a long history and take on some feature of central place patterns. The top class of the hierarchy of human settlements is a provincial capital, Zhengzhou, and the bottom class is composed of various county towns. The total number of urban places is 220. However, only 216 urban places can be identified by the software of geographical information systems (GIS); therefore, we have a size of sample $n$=216.

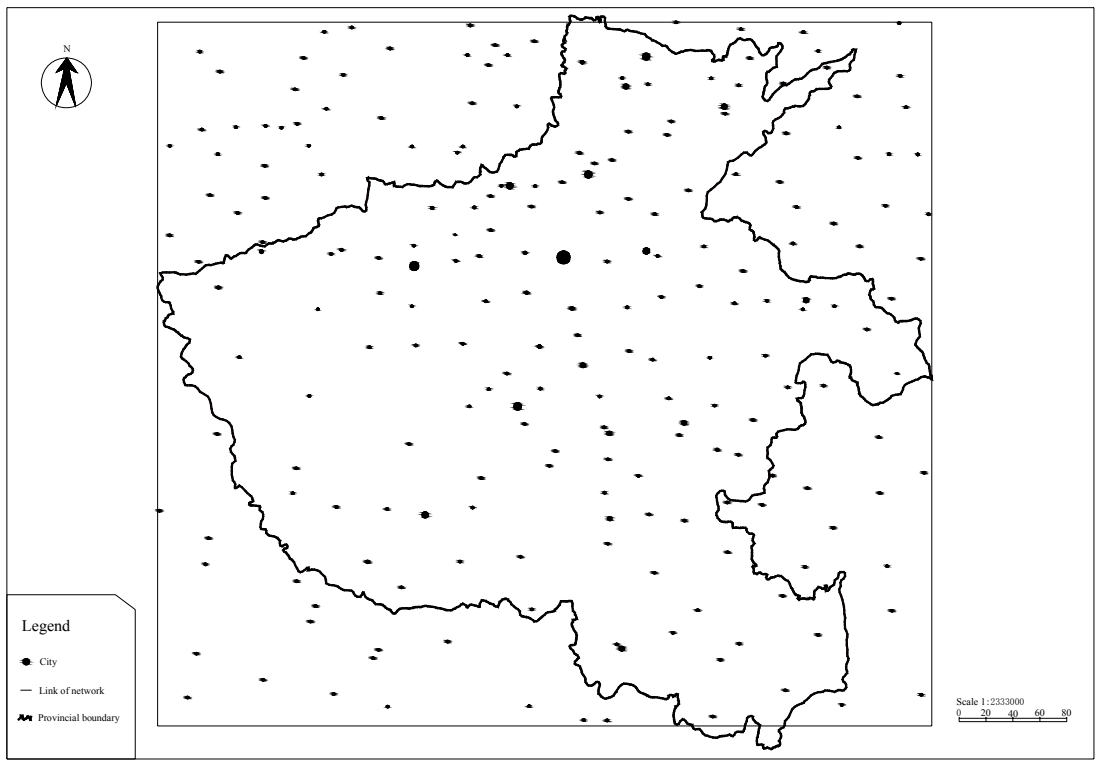

Figure 6 The pattern of the systems of urban places in the Central Plains of China

In this study, the probability measurement will be determined with the functional box-counting method (Chen and Wang, 2013; Feng and Chen, 2010; Lovejoy *et al*; 1987). First step, a rectangular frame is used to cover all the 216 urban places. This is the first-level box with a sidelength of $r$=1 unit. The number of the cities falling into this box is $n$=216. The probability is $P(1)=n/n=216/216=1$, thus the Shannon entropy $I_1(1)=0$, and the Renyi entropy $I_q=0$ ($q\neq 1$). Then, second step, the largest box is equally divided into 4 smaller boxes. These are the second-level boxes, and the side length of each box is $r$=1/2 unit. The numbers of the cities within the four smaller rectangular frames are $n_1=37$, $n_2=45$, $n_3=65$, and $n_4=69$, respectively. Accordingly, the probability measurements are $P_1=37/216$, $P_2=45/216$, $P_3=65/216$, and $P_4=69/216$, respectively. Thus the Shannon entropy is $I_1(1/2)=-P_1\ln(P_1)-P_2\ln(P_2)-P_3\ln(P_3)-P_4\ln(P_4)=-37/216*\ln(37/216)$ $-45/216*\ln(45/216)-65/216*\ln(65/216)-69/216*\ln(69/216)=1.3549$ nat. The Renyi entropy can be given by

$$I_q(1/2) = \frac{1}{1-q}\ln((\frac{37}{216})^q + (\frac{37}{216})^q + (\frac{37}{216})^q + (\frac{37}{216})^q).$$

For example, if $q=0$, we will have $I_0(1/2)=\ln(4)\approx1.3863$; if $q=2$, we will have $I_2(1/2)\approx1.3267$; if $q=-2$, we will have $I_{-2}(1/2)\approx1.4521$. The rest may be derived by analogy. Third step, the 4 second-level boxes are divided into 16 smaller boxes. These are the third-level boxes with a side length of $r=1/4$ unit. Accordingly, we have Shannon entropy and a set of Renyi entropy.

The rest steps can be implemented by analogous procedure. For the $m$th step, the linear size of boxes is $r=1/2^{m-1}$, the total number of boxes is $4^{m-1}$, and the number of nonempty boxes $N(r)\leq 4^{m-1}$ ($m=1,2,3,\ldots$). The Renyi entropy and Shannon entropy can be calculated with equations (21) and (22). If the side length of boxes is very small, the number of nonempty boxes will approach the number of urban places, that is, $N(r)\to n$. In this instance, the information entropy cannot increase significantly if the boxes become smaller. For the case of China's cities, in the 6th step, $r=1/2^5$, we have $N(r)=201$, which is close to $n$; in the 10th step, $r=1/2^9$, we have $N(r)=216=n$, and the information entropy reaches the maximum entropy $I_{max}=\ln(216)$ nat. In this instance, it is meaningless to divide the 10th level boxes into the 11th level boxes. By means of the probability measurements, we can calculate the Renyi information $I_q(r)$ and the two measurements based on the $\mu$-weight, $\sum\mu\ln(P)$ and $\sum\mu\ln(\mu)$ (Table 6).

**Table 6 Partial results of the information entropy and the measurements based on $\mu$-weight**

| Range | $m$ | $r$ | Information entropy $I_q(r)$ | | | $-\sum\mu(r)\ln(P(r))$ | | | $-\sum\mu(r)\ln(\mu(r))$ | | |
|---|---|---|---|---|---|---|---|---|---|---|---|
| | | | $q=0$ | $q=1$ | $q=2$ | $q=0$ | $q=1$ | $q=2$ | $q=0$ | $q=1$ | $q=2$ |
| Scaling range | 1 | $1/2^0$ | 0.0000 | 0.0000 | 0.0000 | 0.0000 | 0.0000 | 0.0000 | 0.0000 | 0.0000 | 0.0000 |
| | 2 | $1/2^1$ | 1.3863 | 1.3549 | 1.3267 | 1.4188 | 1.3549 | 1.3004 | 1.3863 | 1.3549 | 1.2740 |
| | 3 | $1/2^2$ | 2.7726 | 2.7240 | 2.6822 | 2.8239 | 2.7240 | 2.6441 | 2.7726 | 2.7240 | 2.6061 |
| | 4 | $1/2^3$ | 4.1431 | 4.0224 | 3.9305 | 4.2796 | 4.0224 | 3.8505 | 4.1431 | 4.0224 | 3.7706 |
| | 5 | $1/2^4$ | 5.0626 | 4.9772 | 4.8813 | 5.1398 | 4.9772 | 4.7837 | 5.0626 | 4.9772 | 4.6862 |

| | | | | | | | | | | |
|---|---|---|---|---|---|---|---|---|---|---|
| Linear | 6 | $1/2^5$ | 5.3033 | 5.2790 | 5.2452 | 5.3236 | 5.2790 | 5.2062 | 5.3033 | 5.2790 | 5.1672 |
| range | 7 | $1/2^6$ | 5.3471 | 5.3368 | 5.3212 | 5.3555 | 5.3368 | 5.3023 | 5.3471 | 5.3368 | 5.2834 |
| | 8 | $1/2^7$ | 5.3613 | 5.3560 | 5.3479 | 5.3655 | 5.3560 | 5.3378 | 5.3613 | 5.3560 | 5.3277 |
| | 9 | $1/2^8$ | 5.3706 | 5.3689 | 5.3661 | 5.3721 | 5.3689 | 5.3626 | 5.3706 | 5.3689 | 5.3591 |
| | 10 | $1/2^9$ | 5.3753 | 5.3753 | 5.3753 | 5.3753 | 5.3753 | 5.3753 | 5.3753 | 5.3753 | 5.3753 |

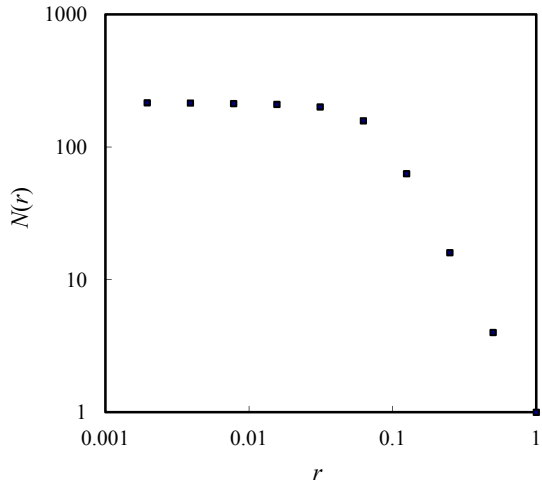
a. Scatter of points

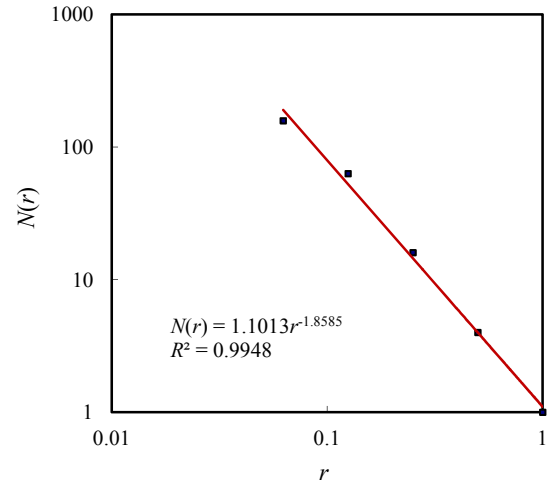
b. Capacity dimension

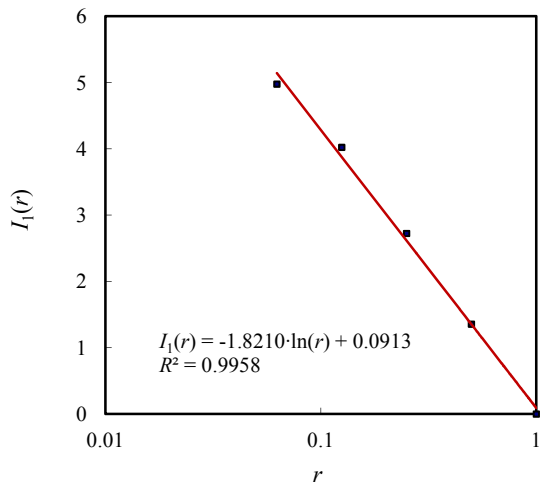
c. Information dimension

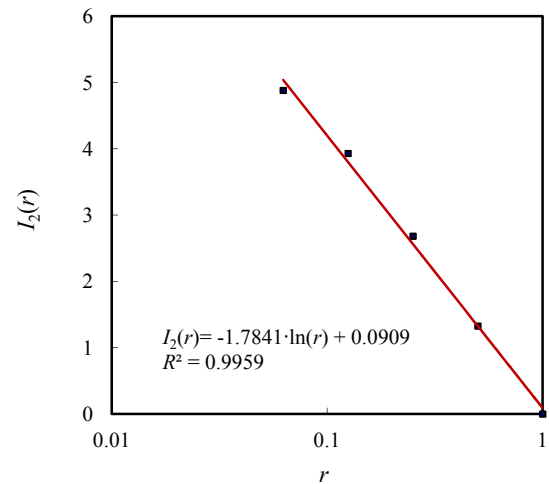
d. Correlation dimension

**Figure 7 A set of plots showing fractal dimension estimation**

(**Note**: The first plot displays the double logarithmic relationships between the linear size of boxes and the number of nonempty boxes, and the other plots show the scaling ranges for the capacity dimension, information dimension, and correlation dimension.)

As the number of cities is limited, the scaling relation between the linear scale and the corresponding measurements based on the probability is also confined to certain range. In other words, not all the observational data can be used to estimate fractal parameters. If we show the double logarithmic relationships between the linear size of boxes $r$ and the number of nonempty boxes $N(r)$ using a log-log plot, we can find that the data points can be fitted into two line segments (Figure 7(a)). The first 5 levels of data points ($m$=1~5) can be fitted to the first line segment, which represents a scaling range, and its slope gives the capacity dimension (Figure 7(b)). The last 5 levels of data points ($m$=6~10) satisfy a linear relation rather than a power law relation, and they form a quasi-horizontal line on the log-log plot (Figure 7(a)). The semi-logarithmic relationships between the information entropy $I_q(r)$ and the corresponding linear scale $r$ can also be fitted to two line segments, and the first 5 levels of data points that can be fitted into equation (20) can yield the fractal parameters $D_q$ and $\tau_q$ (Figure 7(c), Figure 7(d)). Similarly, we can estimate the local parameters $\alpha(q)$ and $f(\alpha)$ by means of the linear relationships between $\ln(r)$ and $\sum\mu(r)\ln(P(r))$ as well as $r$ and $\sum\mu(r)\ln(\mu(r))$. Parts of the values of the global and local parameters yielded by linear regression analyses are tabulated as follows (Table 7). Using these results, we can obtain the spectral curves of multifractal parameters (Figure 8); moreover, we can draw the unimodal curve of the local dimension changing along with the singularity exponent (Figure 8).

Table 7 Partial values of global and local parameters of multifractal structure of the system of cities in Central Plains of China

| $q$ | Global parameter | | | | Local parameter | | | |
|---|---|---|---|---|---|---|---|---|
|  | $D_q$ | $R^2$ | $\tau_q$ | $R^2$ | $\alpha(q)$ | $R^2$ | $f(\alpha)$ | $R^2$ |
| -100 | 2.0676 | 0.9477 | -208.8276 | 0.9477 | 2.0719 | 0.9454 | 1.6370 | 0.7553 |
| -50 | 2.0634 | 0.9499 | -105.2334 | 0.9499 | 2.0719 | 0.9454 | 1.6369 | 0.7564 |
| -20 | 2.0511 | 0.9559 | -43.0731 | 0.9559 | 2.0725 | 0.9456 | 1.6231 | 0.7875 |
| -10 | 2.0306 | 0.9642 | -22.3366 | 0.9642 | 2.0755 | 0.9469 | 1.5819 | 0.8538 |
| -5 | 1.9925 | 0.9757 | -11.9550 | 0.9757 | 2.0746 | 0.9518 | 1.5819 | 0.9249 |
| -2 | 1.9280 | 0.9885 | -5.7840 | 0.9885 | 2.0196 | 0.9725 | 1.7448 | 0.9973 |
| -1 | 1.8949 | 0.9924 | -3.7898 | 0.9924 | 1.9653 | 0.9847 | 1.8245 | 0.9973 |
| 0 | 1.8585 | 0.9948 | -1.8585 | 0.9948 | 1.8958 | 0.9930 | 1.8585 | 0.9948 |
| 1 | 1.8210 | 0.9958 | 0.0000 | 0.9958 | 1.8210 | 0.9958 | 1.8210 | 0.9958 |
| 2 | 1.7841 | 0.9959 | 1.7841 | 0.9959 | 1.7482 | 0.9956 | 1.7123 | 0.9954 |
| 5 | 1.6940 | 0.9943 | 6.7760 | 0.9943 | 1.5986 | 0.9914 | 1.2174 | 0.9637 |
| 10 | 1.6092 | 0.9917 | 14.4828 | 0.9917 | 1.5024 | 0.9869 | 0.5416 | 0.6064 |
| 20 | 1.5393 | 0.9888 | 29.2467 | 0.9888 | 1.4646 | 0.9850 | 0.0444 | 0.0061 |
| 50 | 1.4918 | 0.9869 | 73.0982 | 0.9869 | 1.4615 | 0.9859 | 0.0250 | 0.0037 |
| 100 | 1.4766 | 0.9865 | 146.1834 | 0.9865 | 1.4618 | 0.9861 | 0.0026 | 0.0001 |

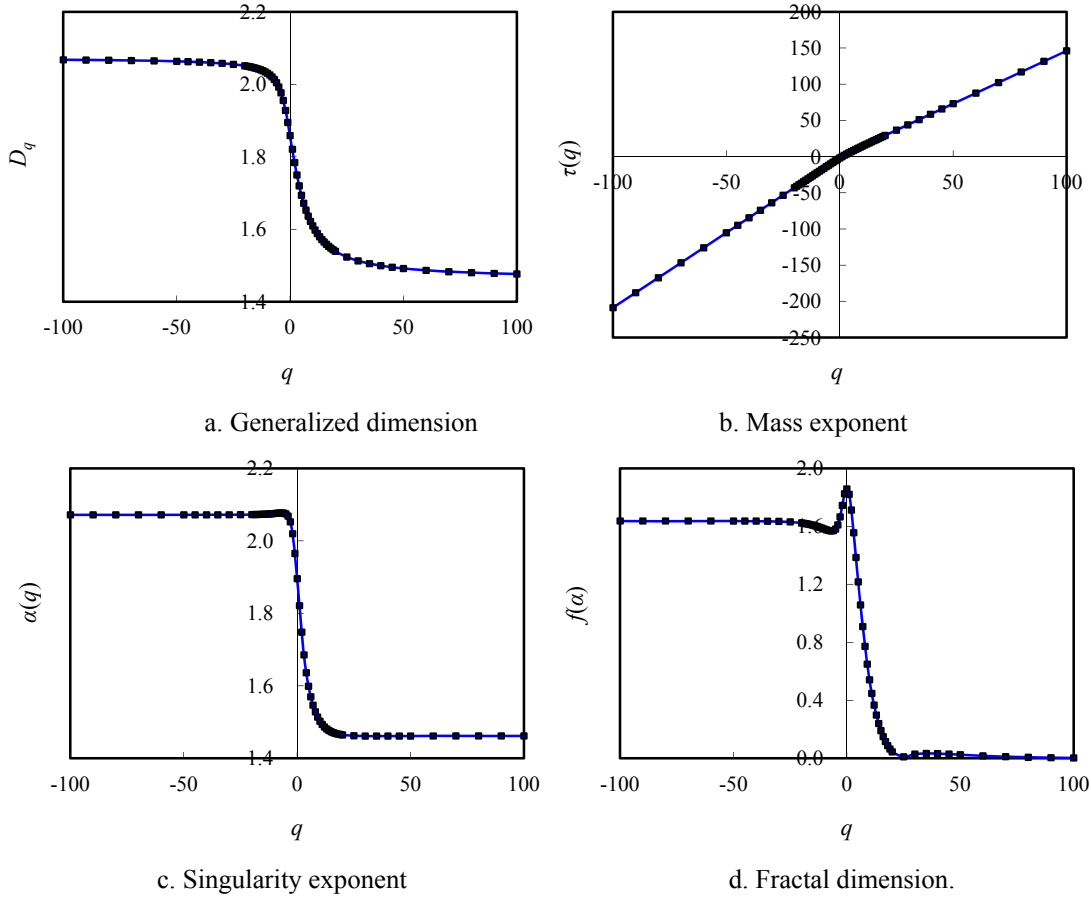

a. Generalized dimension      b. Mass exponent

c. Singularity exponent      d. Fractal dimension.

**Figure 8 The singularity spectrums of multifractal structure of the system of cities in Central Plains of China**

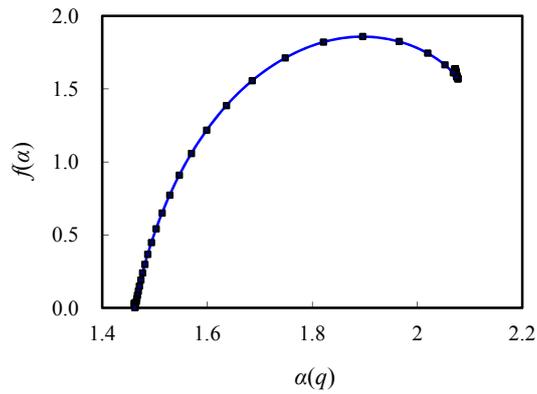

**Figure 9 The *f(α)* curves of the multifractal structure of the systems of cities in the Central Plain, China**

## 3.3 Multifractal analysis of urban systems

Before making a spatial analysis of the multifractal structure of the systems of urban places in the Central Plain, two concepts of analogy must be explained. First, the order of moment, $q$, can be assimilated to a zoom lens and a wide-angle lens. Changing the $q$ value, we can examine the coarse patterns and fine structures from different levels and angles of views. Second, $q$ can be assimilated to an adjustor of spatial observation and analysis. If the $q$ value changes from $-\infty$ to $\infty$, the spatial structure from the central part to the peripheral parts of a fractal system will be intensified or brought into focus step by step. Moreover, different areas of probabilities of growth are to a growing multifractal object what different areas of distribution densities are to a city or system of cities. Higher probability of fractal growth corresponds to the higher urban density (the central part of Figure 1, the peripheral parts of Figure 2), while lower probability of fractal growth corresponds to the lower density of city development (the peripheral parts of Figure 1, the central part of Figure 2).

A multifractal object has more than one fractal dimension, that is, different fractal copies may possess different fractal dimension. This differs from a monofractal. A simple fractal has an exclusive dimension: the fractal dimension of one part equals that of another part. If we compute the dimension of each part of a multifractal object, the procedure will be complicated. A practical approach is to utilize the parameter value of the moment order, $q$. By equation (23), the effect of changing $q$ value is as follows. If $q=1$, a fractal pattern has no distortion, and we can examine the real pattern and structure of a fractal. If $q>1$ and $q\to\infty$, the areas of higher growing probabilities (higher density) will be gradually magnified, while the areas of lower growing probabilities (lower density) will be minified step by step. The magnification or minification of a fractal copy is in (direct) proportion to its probability of growth (distribution density). If $q<1$ and $q\to-\infty$, the areas of higher growing probabilities (higher density) will be gradually minified, while the areas of lower growing probabilities (lower density) will be magnified step by step. Thus the larger fractal copies and the smaller copies will be reversed. The magnification or minification of a fractal copy is in inverse proportion to its growing probability (distribution density). For a monofractal object with equal probability of growth, the pattern cannot be transformed by changing $q$ value. Especially, if $q=0$, the probabilities of growth in different parts are regarded as equal. Altering $q$ value is a process of increasing coarse-graining level of a fractal object. Different $q$ values bring different parts into prominence in terms of certain probabilities of growth.

Using the global parameters, we can examine the macrostructure of a multifractal system. The mass exponent $\tau(q)$ reflects the extent of magnification of the fractal object. The higher the $q$ value is, the more the parts of high probability of growth (high density) are zoomed in; the lower the $q$ value is, the more the parts of low probability of growth (low density) are blown up. Consequently, if the parts of high probability of growth (high density) are enlarged, the absolute value of the mass exponent will be high. The generalized correlation dimension $D_q$ reflects the spatial autocorrelation, similarity, and space filling. If $q>1$ and $q\rightarrow\infty$, the fractal parts of high probability of growth (high density) will be magnified, and thus the spatial difference of the fractal pattern becomes bigger and bigger, while the displacement of spatial correlation becomes shorter and shorter. As a result, the $D_q$ value will become lower and lower till its lower limit $D_\infty$. Contrarily, if $q<1$ and $q\rightarrow-\infty$, the fractal parts of low probability of growth (low density) are will be enlarged, and thus the spatial difference of the fractal pattern becomes smaller and smaller, while the displacement of spatial correlation becomes longer and longer. In this instance, the $D_q$ value will become higher and higher up to its upper limit $D_{-\infty}$.

Using the local parameters, we can examine the microstructure of a multifractal system. The singularity exponent $\alpha(q)$ indicates the parts that can be reflected by a local dimension $f(\alpha)$. Different parts of a multifractal object will be brought into focus step by step by changing the $q$ value. For the $i$th part with a growing probability $p_i$ and linear scale $r_i$, if and only if the relationships between $p_i$ and $r_i$ satisfies equation (3), i.e., $p_i=r_i^{\alpha(q)}$, the fractal dimension of this part will be given by equation (4) and we will have $f(\alpha)$. At the same time, the fractal dimension of other parts will be neglected unless they satisfy the condition $p_j=r_j^{\alpha(q)}$ ($j\neq i$). As a result, the local dimension $f(\alpha)$ based on different $q$ values corresponds to different parts of the fractal object. This differs from the global dimension $D_q$, which reflects the spatial character of the whole fractal pattern. If $q<0$ and $q\rightarrow-\infty$, the fractal parts with low probability of growth (low density) will be focalized, and the corresponding local dimensions will come forth. When $q$ approaches minus infinity, we have the fractal dimension values of the peripheral areas of the concentration pattern or those of central areas of the deconcentration pattern. Contrarily, if $q>1$ and $q\rightarrow\infty$, the fractal parts with high growing probability (high density) will be in focus, and the corresponding local dimensions will manifest. When *q* approaches infinity, we have the fractal dimension values of the central areas of the concentration pattern or those of peripheral areas of the deconcentration pattern.

By means of the ideas from multifractals, the main geographical feature of the system of cities and towns in the Central Plain can be revealed as follows.

**First, the spatial pattern of geographical distribution of urban places reflects the concentration process of urban evolution**. Comparing Figure 8 with Figures 3 and 4 shows that the forms of Figure 8 is similar to Figure 3 rather than Figure 4, and comparing Figure 9 with Figure 5 shows that the form of Figure 9 is similar to Figure 5(a) instead of Figure 5(b). In other words, the multifractal parameter spectrums of the system of human settlements in the study area bear an analogy to those of the multifractal model of spatial concentration. The difference between the systems of urban places and the first multifractal model of central places is that the global dimension values are greater than the corresponding parameter values of the fractal model. This suggests that it is a "centripetal force" or "pull of center" rather than a "centrifugal force" or "push of center" that played a more important role in the process of spatial development of the urban systems.

**Second, the multifractal structure of the systems of cities was well developed at the macro level**. The valid range of the moment order *q* is wide and it varies from -131 to 177. When *q*=-131(peripheral part), the generalized correlation dimension reach its upper limit, and we have $D_{-131}=2.0686$, the corresponding goodness of fit is $R^2=0.9471$; When *q*=177 (central part), the global dimension reach its lower limit, and we have $D_{177}=1.4701$, the corresponding determination coefficient is around $R^2=0.9863$. If *q*<-131 or *q*>177, the scaling relationship between the linear size of boxes and the Renyi entropy will be broken and diverge from the linear regression. From *q*=-131 to *q*=177, the values of generalized correlation dimension $D_q$ vs *q* form an inverse sigmoid curve (Figure 8(a)). Accordingly, the values of mass exponent $\tau_q$ vs *q* form two line segments (Figure 8(b)). The slopes of the line segments give the estimated values of the upper and lower limits of $D_q$, respectively, that is $D_{-131}=2.0718$ and $D_{177}=1.4700$.

**Third, the multifractal structures of the subsystem of cities at the micro level were not very developed**. Despite the fact the valid *q* range of the singularity exponent $\alpha(q)$ is the same as that of the global dimension $D_q$, the valid *q* range of the local dimension is very narrow (Figure 8(c-d)). If $q<$-7 (peripheral parts), the local scaling relations is not well developed; if $q>7$ (central parts), the local scaling relations go rapidly into breakdown, and this can be reflected by the goodness of fit, $R^2$ (Table 7). From the centers of subsystems to the main center of the urban system, the local scaling relations were influenced by some negative factors such as improper city planning and unsuited development. From the peripheral areas of subsystems to the peripheral area of the urban system, the scaling relations were gradually broken.

## 4. Questions and discussion

The central place multifractals presented in Section 2 are theoretical works which can only appears in the mathematical world rather than the real world. However, they are useful for understanding geographical multifractals and making spatial scaling analysis of urban evolution. The models are regular, and the spectrum curves of multifractal parameters are typical, so it is easy to grasp the spatial implications. In practice, we can draw a comparison between the empirical multifractal dimension spectrums of urban systems with the theoretical multifractal dimension spectrums of mathematical models. If the singularity spectrums of actual systems of human settlements are similar to those displayed in Figure 2 and Figure 5(a), the geographical process will be of spatial concentration; while if the singularity spectrums of real urban systems bear an analogy to those displayed in Figure 4 and Figure 5(b), the geographical process will be of spatial deconcentration. As indicated above, the multifractal parameter spectrums of the systems of urban places in the Central Plain displayed in Figures 8 and 9 are similar to those shown in Figure 2 and Figure 5(a). This indicates that it was the concentration process rather than the deconcentration process that dominated the urban evolution of the study area.

The theoretical foundation of fractal central places differs from that of traditional central place concepts. The central place theory of Christaller (1933/1966) is based on a series of assumptions of spatial conditions: an unbounded isotropic, homogeneous, limitless surface (all flat), an evenly distributed population and resources, distance decay mechanism, and so on. From this kind of postulates, Christaller (1933/1966) deduced that human settlements would tend to form in a hexagonal lattice, and all the settlements are equidistantly distributed in the triangular lattice pattern. This used to be regarded as the most efficient pattern to serve areas without any overlap.

However, an actual system of central places is a self-organized system and may be far from equilibrium (Allen, 1997). In the real world, the regular hierarchical distribution is very infrequent because historical, political, and geographical, and political factors abound and disrupt this kind of spatial symmetry (Prigogine and Stengers, 1984). In fact, the classical central place theory is based on the concept of complete space-filling with no intermittency. The theory predicted a Euclidean space with dimension $d$=2. However, this result is not consistent with the empirical values based on observational data of systems of urban places (Chen and Zhou, 2006). Introducing the ideas of incomplete space-filling and spatial intermittency into urban studies, we can construct the monofractal models of central places (Chen, 2011). Further, introducing the concept of inequiprobable growth into spatial analysis, we can make multifractal models of central place snowflakes.

The spatial intermittency is associated with the shadow effect of urban evolution. The *shadow effect* suggests that it is hard for smaller cities and towns to grow and develop in the neighborhood (adjacent area) of a large city (Chen, 2011; Chen and Zhou, 2006). In fact, a large city cast it "shadow" over the surrounding country depriving the smaller cities and towns of growth as a larger tree prevents the growth of others by depriving them of light (Evans, 1985). Simple fractal models of central places fall into two types: one is with no shadow effect, and the other with shadow effect. Multifractal central place models also fall into two classes: one is with no shadow effect (Figure 1), and the other with shadow effect (Figure 3). All these models are based on the incomplete space-filling process.

Now, the fractal systems of central places can be divided into two types: one is the simple system which can be described with monofractals, and the other is the complex systems which should be characterized with mulitfractals. The traditional approach to identifying a central place system is to draw an analogy between a network of human settlements and a triangular lattice pattern. However, the conclusions are always plausible or even specious. The essence of central place systems is geographical scaling rather than regular hexagon. Based on the ideas from fractal and symmetry, a central place system can be identified by scaling laws and coordination law of city distribution. The scaling laws say that the size, number and distance of the urban places in central place system take on scale invariance. Central place scaling consists of spatial scaling, hierarchical scaling, and allometric scaling (Chen, 2011). The coordination law says that average number of urban places surrounding an urban place approaches to 6, which is termed coordination number (Ye *et al*, 2001). The 6 coordination cities or towns can be linked to form a convex polygon. Further, if we want to identify the singularity and complex structure of a central place system, we will have to rely heavily on the multifractal parameter spectrums.

The shortcoming of this study is as follows. In theory, only two models are proposed, and the two models cannot reflect all patterns of spatial distributions of cities in the real world. In the empirical analysis, only the case of spatial concentration was taken into consideration owing to limit of space of an article. What is more, the small towns were not taken into account in the sample due to absence of complete spatial data set in the digital map. The case of spatial deconcentration was supported by the multifractal form of Beijing city, China (Chen and Wang, 2013). The theoretical models will be further developed, and better empirical studies will be presented in the future.

## 5. Conclusions

This is a theoretical study on complex systems of central places, which can be described with fractal geometry. Using the ideas from growing multifractals, I construct two models of fractal central place systems. One is used to reflect the process of spatial concentration, and the other is to mirror the process of spatial deconcentration. Based on the typical models, two sets of typical multifractal dimension spectrums are presented. In light of the typical dimension spectrums, we can analyze the real geographical process of urban evolution. The modeling results form a simple approach to understanding two types of geographical patterns and the corresponding processes. In particular, the models provide a referential framework for multifractal analysis of real systems of urban places.

Based on the theoretical modeling, empirical analysis and discussion of questions, the main conclusions of this paper can be drawn as follows.

**First, multifractal geometry can be employed to model the complex systems of central places.** Central place fractals fall into two types: one is monofractals of central places, which suggest equal probability of growth, and the other is multifractals of central places, which indicate unequal probabilities of growth of spatial evolution. If different parts of a real system of cities and towns have different chances of development, the system may be complex enough to contain multi-scaling processes. In this instance, the system of urban places should be described with the multifractal method because the monofractal models are not enough to reflect its complexity and singularity.

**Second, multifractal systems of central places may have different mechanisms of evolution.** Central place multifractals can be divided into two typical types: one is based on the geographical process of spatial concentration, and the other is based on the process of spatial deconcentration. The former seems to be controlled by geographical centripetal force or gravitational force of centers, and the other seems to be dominated by geographical centrifugal force or repulsion force of central cities. Different types of central place multifractals have different singularity spectrums, which in turn provide a criterion to identify different geographical processes of urban evolution.

**Third, multifractal parameter spectrums provide an important approach to analyzing the central place systems in the real world.** A real system of central places is more complex than the theoretical models. However, it can be researched by using multifractal dimensions. The analytical procedure consists of the following steps. Step 1: whether or not a systems of urban places is multifractals? This can be judged by the capacity dimension, information dimension, and correlation dimension. Step 2: if the system is multifractals, what type of central place multifractals it belongs to? This can be identified by drawing an analogy between the singularity spectrums of the real system and those of the theoretical models. Step 3: whether or not there exist some problems in the process of urban evolution? The macro-level problems can be revealed through the global parameter spectrums, while the micro-level problems can be disclosed with the local parameter spectrums.

## Acknowledgment

This research was sponsored by the National Natural Science Foundation of China (Grant No. 41171129). The supports are gratefully acknowledged.